\documentclass[aps,preprint]{revtex4}%
\usepackage{amsfonts}
\usepackage{amsmath}
\usepackage{amssymb}
\usepackage{graphicx}%
\begin{document}
\preprint{ }
\title[ ]{The Effect of Strange Quark on the Chiral Symmetry Breaking in Magnetic
Background in the framework of logarithmic quark-sigma model}
\author{M. Abu-Shady}
\affiliation{{\small Department of Applied Mathematics, Faculty of Science, Menoufia
University, Egypt}}
\author{}
\affiliation{}
\author{}
\affiliation{}
\affiliation{}
\pacs{PACS number}

\begin{abstract}
The chiral symmetry breaking in the presence of an external magnetic field is
studied in the framework of logarithmic quark-sigma model with three flavors
$\left(  u,d,s\right)  $. The effective logarithmic mesonic potential is
extended to strange quark and is numerically solved in the mean-field
approximation. The present results show that the chiral symmetry breaking is
sensitive for the strange quark flavour in the presence an external magnetic
field. In addition, the effect of the free parameters is studied. A comparison
with original sigma model is discussed.


\end{abstract}
\volumeyear{ }
\volumenumber{ }
\issuenumber{ }
\eid{ }
\startpage{1}
\endpage{ }
\maketitle

\section{Introduction}

The quantum chromodyanmic (QCD) theory is an acceptable theory for strong
interactions. A direct technique to calculate the hadron properties through
using the lattice QCD but this is not easy task, in particular, at finite
density. So the investigations are preformed by using effective models such as
the Nambu-Jona-Lasinio model and quark sigma model that share the QCD theory
in same properties. The linear sigma model is introduced by Gell-Man and Levy
$\left[  1\right]  $ to describe pion-nucleon interactions. This model is
extended to quark level by Birse and Banerjee $\left[  2\right]  $ to
calculate hadron properties at low energy. Birse and Banerjee model $\left[
2\right]  $ has some difficulties that conflict with experimental data. So,
the model is extended to avoid these difficulties as in Refs. $\left[
3-5\right]  $. In addition, the model is extended to include finite
temperature as Refs. $\left[  6-8\right]  $.

The study of the influence of external magnetic fields on the fundamental
properties of quantum chromodynamic (QCD) theory, confinement and dynamical
chiral symmetry breaking is still\ a matter of great interest theoretical and
experimental activities. In Ref. $\left[  9\right]  $, the fact that an
external magnetic field enhances the generation of a fermion mass in 3+1
dimensions was first established in the framework of the Nambu-Jona-Lasinio
(NJL) model made in Refs. $\left[  10,11\right]  $. Also, in the context of
the (2+1)-dimensional Gross-Neveu model, which was expected to give a simple
effective description for certain condensed matter planar systems, the authors
of Refs. $\left[  12,13\right]  $ showed that the dynamical generation of
nonzero fermion mass takes place in a magnetic field as soon a there an
attractive interaction between fermions and antifermions. The symmetric quark
matter and quark matter in $\beta$ equilibrium are investigated in the NJL
model $\left[  14\right]  $ in the presence external magnetic field at zero
temperature $\left[  15,16\right]  $. The effective quark model that takes
into account chiral symmetry $\left[  17,18\right]  $. It was shown that
non-negligible effects on the equation of state and single-particle quark
properties. The linear sigma model with two flavors has also applied to
determine the critical point temperature in the presence of external magnetic
field $\left[  19,20\right]  $. The effect of the higher-order mesonic
interactions on the chiral symmetry breaking is investigated in the presence
of an external magnetic field $\left[  21\right]  $.

In Ref. $\left[  22\right]  ,$ the authors have modified the linear sigma
model by including the logarithmic mesonic potential and study its effect on
the phase transition at finite temperature. In addition, the comparison with
other models is done. On the same hand, the logarithmic sigma model
successfully to describe the nucleon properties at finite temperature and
density $\left[  23\right]  $. In addition, the logarithmic sigma model give
good descriptions of magnetic catalysis for two flavors $\left[  24\right]  $.

To continue the investigation that started in Ref. $\left[  22\right]  $, the
effect of strange quark field on the chiral symmetry breaking in the framework
of logarithmic sigma model is investigated in the presence of an external
magnetic field. So for no attempts have done to investigate the logarithmic
sigma model for three flavors $\left(  u,d,s\right)  $ in the presence of an
external magnetic field. In addition, a new parametrization set for sigma mass
and coupling constant on the behavior of chiral symmetry breaking is studied.

This paper is organized as follows: The background of previous papers is given
in Sec. 1. The original sigma model is briefly presented in Sec. 2. Next, the
effective logarithmic mesonic potential in the presence of external magnetic
field is presented in Sec. 3. The results are discussed and are compared with
other models in Secs. 4 and 5, respectively. Finally, the summary and
conclusion are presented in Sec. 6.

\section{The chiral sigma model with original effective potential}

The interactions of quarks via the exchange of $\sigma-$ and $\mathbf{\pi}$
-meson fields are given by the Lagrangian density $[2]$ as follows:
\begin{equation}
{\small L}\left(  r\right)  {\small =i}\overline{\Psi}{\small \partial}_{\mu
}{\small \gamma}^{\mu}{\small \Psi+}\frac{1}{2}\left(  \partial_{\mu}%
\sigma\partial^{\mu}\sigma+\partial_{\mu}\mathbf{\pi}.\partial^{\mu
}\mathbf{\pi}\right)  {\small +g}\overline{\Psi}\left(  \sigma+i\gamma
_{5}\mathbf{\tau}.\mathbf{\pi}\right)  {\small \Psi-U}_{1}{\small (\sigma
,\pi),} \tag{1}%
\end{equation}

with
\begin{equation}
{\small U}_{1}{\small (\sigma,\pi)=}\frac{\lambda^{2}}{4}\left(  \sigma
^{2}+\mathbf{\pi}^{2}-\nu^{2}\right)  ^{2}+{\small m}_{\pi}^{2}{\small f}%
_{\pi}{\small \sigma.}\tag{2}%
\end{equation}
$U_{1}\left(  \sigma,\mathbf{\pi}\right)  $ is the meson-meson interaction
potential where $\Psi,\sigma$ and $\mathbf{\pi}$ are the quark, sigma, and
pion fields, respectively. In the mean-field approximation, the meson fields
are treated as time-independent classical fields. This means that we replace
the power and products of the meson fields by corresponding powers and the
products of their expectation values. The meson-meson interactions in Eq. (2)
lead to hidden chiral $SU(2)\times SU(2)$ symmetry with $\sigma\left(
r\right)  $ taking on a vacuum expectation value \ \ \ \ \ \ \
\begin{equation}
\ \ \ \ {\small \ \ }\left\langle \sigma\right\rangle {\small =-f}_{\pi
}{\small ,}\tag{3}%
\end{equation}
where $f_{\pi}=93$ MeV is the pion decay constant. In Eq. (2), the final\ term
is included to break the chiral symmetry explicitly. It leads to the partial
conservation of axial-vector isospin current (PCAC). The parameters
$\lambda^{2}$and $\nu^{2}$ can be expressed in terms of$\ f_{\pi}$, sigma and
pion masses as,
\begin{equation}
{\small \lambda}^{2}{\small =}\frac{m_{\sigma}^{2}-m_{\pi}^{2}}{2f_{\pi}^{2}%
}{\small ,}\tag{4}%
\end{equation}%
\begin{equation}
{\small \nu}^{2}{\small =f}_{\pi}^{2}{\small -}\frac{m_{\pi}^{2}}{\lambda^{2}%
}{\small .}\tag{5}%
\end{equation}

\section{The effective logarithmic potential in the presence of magnetic
field}

In this section, the logarithmic mesonic potential $U_{2}(\sigma,\pi)$ is
employed. In Eq. (6), The effective logarithmic potential is extended to
include the external magnetic field at zero temperature and chemical potential
as follows,%

\begin{equation}
U_{eff}(\sigma,\pi)=U_{2}(\sigma,\pi)+U_{Vaccum}+U_{Matter},\tag{6}%
\end{equation}
where%
\begin{equation}
\ \ \ \ \ \ U_{2}\left(  \sigma,\mathbf{\pi}\right)  =-\lambda_{1}^{2}\left(
\sigma^{2}+\mathbf{\pi}^{2}\right)  +\lambda_{2}^{2}\left(  \sigma
^{2}+\mathbf{\pi}^{2}\right)  ^{2}\log\left(  \frac{\sigma^{2}+\mathbf{\pi
}^{2}}{f_{\pi}^{2}}\right)  +m_{\pi}^{2}f_{\pi}\sigma,\tag{7}%
\end{equation}
In Eq. 7, the logarithmic potential satisfies the chiral symmetry when
$m_{\pi}\longrightarrow0$ as well as in the standard potential in Eq. 2.
Spontaneous chiral symmetry breaking gives a nonzero vacuum expectation for
$\sigma$ and the explicit chiral symmetry breaking term in Eq. 6 gives the
pion its mass.%
\begin{equation}
\left\langle \sigma\right\rangle =-f_{\pi}.\tag{8}%
\end{equation}
Where%
\begin{equation}
\ \ \ \ \ \lambda_{1}^{2}=\frac{m_{\sigma}^{2}-7m_{\pi}^{2}}{12},\tag{9}%
\end{equation}%
\begin{equation}
\ \ \ \ \ \lambda_{2}^{2}=\frac{m_{\sigma}^{2}-m_{\pi}^{2}}{12f_{\pi}^{2}%
}.\tag{10}%
\end{equation}
For details, see Refs. $\left[  22,23\right]  $. To include the external
magnetic field in the present model, we follow Ref. $\left[  20\right]  $ by
including the pure fermionic vacuum contribution in the free potential energy.
Since this model is renormalizable the usual procedure is to regularize
divergent integrals using dimensional regularization and to subtract the ultra
violet divergences. This procedure gives the following result%
\begin{equation}
{\small U}_{Vaccum}=\frac{N_{c}N_{f}~g^{4}}{(2\pi)^{2}}(\sigma^{2}+\pi
^{2})^{2}(\frac{3}{2}-\ln(\frac{g^{2}(\sigma^{2}+\pi^{2})}{\Lambda^{2}%
})),\tag{11}%
\end{equation}
where $N_{c}=3$ and $N_{f}=3$ are color and flavor degrees of freedom.
respectively, and $\Lambda$ is mass scale,
\begin{equation}
U_{Matter}=\frac{N_{c}}{2\pi^{2}}\sum_{f~=u,d,s}(\left\vert q_{f}\right\vert
B)^{2}[\zeta^{(1,0)}(-1,x_{f})-\frac{1}{2}(x_{f}^{2}-x_{f})\ln x_{f}%
+\frac{x_{f}^{2}}{4}]\tag{12}%
\end{equation}
In Eq. 12, we have used $x_{f}=\frac{g^{2}(\sigma^{2}+\pi^{2})}{(2\left\vert
q_{f}\right\vert B)}$ and $\zeta^{(1,0)}(-1,x_{f})=\frac{d~\zeta(z,x_{f})}%
{dz}\mid_{z~=-1}$ that represents the Riemann-Hurwitz function, and also
$\left\vert q_{f}\right\vert $ is the absolute value of quark electric charge
in external magnetic field with intense $B$ and the potential in Eq. 12 is
extended to include strange quark based on Ref. $\left[  25\right]  $. In Eq.
6, the of effect of the finite temperature and chemical potential is not
included in the present model, in which the present model focus on the study
of magnetic catalysis at low energy due to the enhancement in the chiral
symmetry breaking in the bound state.

\section{ Discussion of results}

In this section, we study the effective potential of logarithmic sigma model.
For this purpose, we numerically calculate the effective potential in Eq. (6).
The parameters of the present model are the coupling constant $\left(
g\right)  $ and the sigma mass $\left(  m_{\sigma}\right)  $. The choice of
free parameters of $\left(  g\right)  $ and $\left(  m_{\sigma}\right)  $
based on Ref. $\left[  20\right]  .$ The parameters are usually chosen so that
the chiral symmetry is spontaneously broken in the vacuum and the expectation
values of the meson fields. In this work, we consider two different sets of
parameters in order to get high and a low value for sigma mass. The first set
is given by $\Lambda=16.48$ MeV which yields $m_{\pi}=138$ MeV and $m_{\sigma
}=600$ MeV. The second set as the first, yielding $m_{\sigma}=400$ MeV.

In Fig. 1, the effective potential is plotted as a function of sigma field for
different values of magnetic field $(B)$. By ignoring\ the $B$ independent one
loop term in Eq. (6). At zero magnetic field, we note that qualitative
agreement between the effect of light quark field and strange quark field. The
difference appears that the effective potential shifts to lower values at
lower values of sigma field. Therefore, we deduce that the spontaneous
symmetry breaking remains unchanged under at zero magnetic field by including
strange quark field. By increasing the external magnetic field as in Fig. (2),
we note that the spontaneous symmetry breaking is clearly appeared and the
potential has largest two minima values in comparison with their values in the
case of light quark. Therefore, the strange quark increases the generated
fermionic mass which leads to increase magnetic catalysis in the present model.%

\begin{center}
\includegraphics[
height=4.0413in,
width=4.0413in
]%
{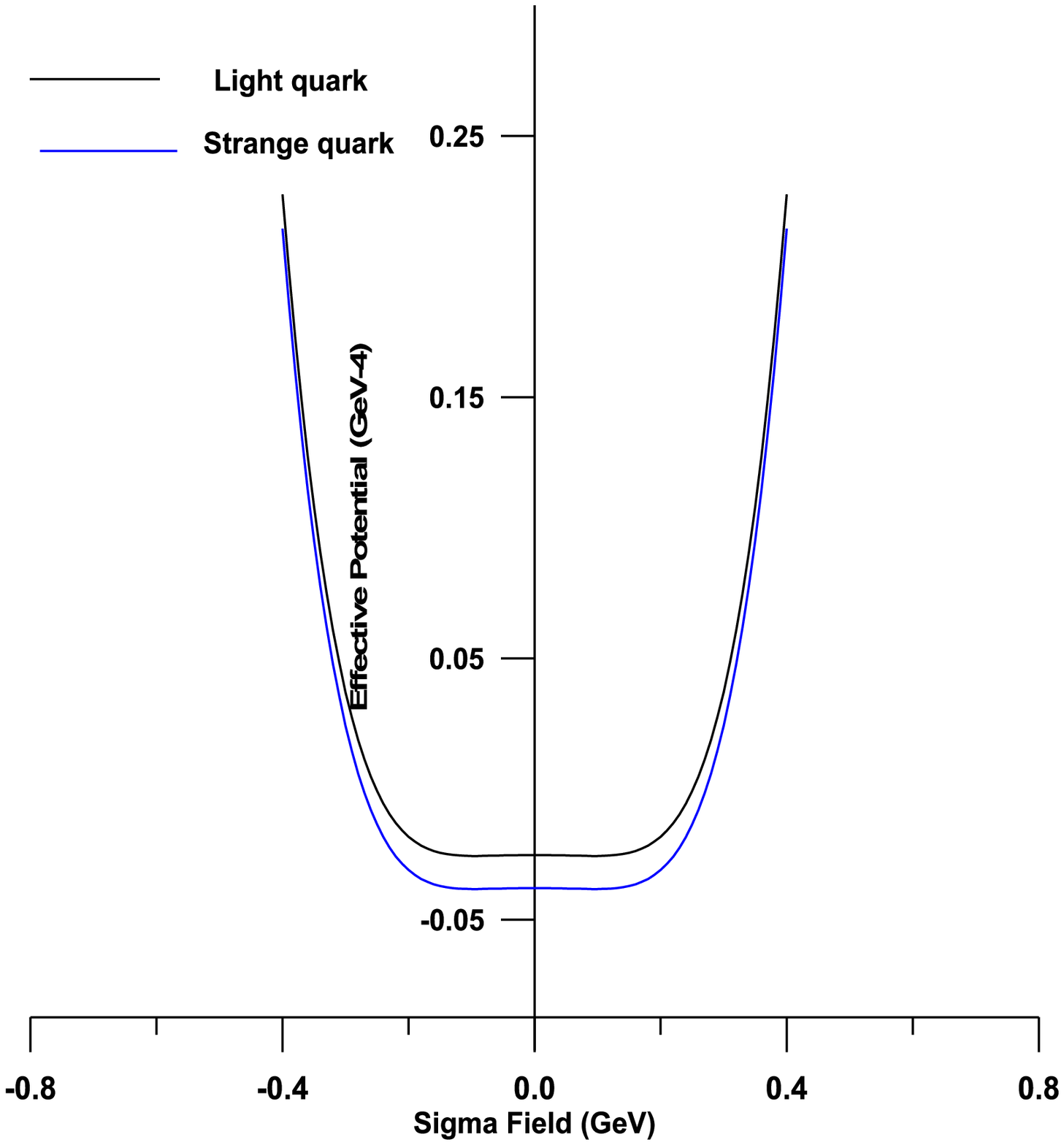}%
\\
\textbf{Fig. 1:} {\small The effective potential is plotted as a function of
sigma field for }$m_{\sigma}=600${\small \ MeV and }$g=4.5${\small \ at zero
magnetic field in the chiral limit}%
\end{center}
%

\begin{center}
\includegraphics[
height=4.0413in,
width=4.0413in
]%
{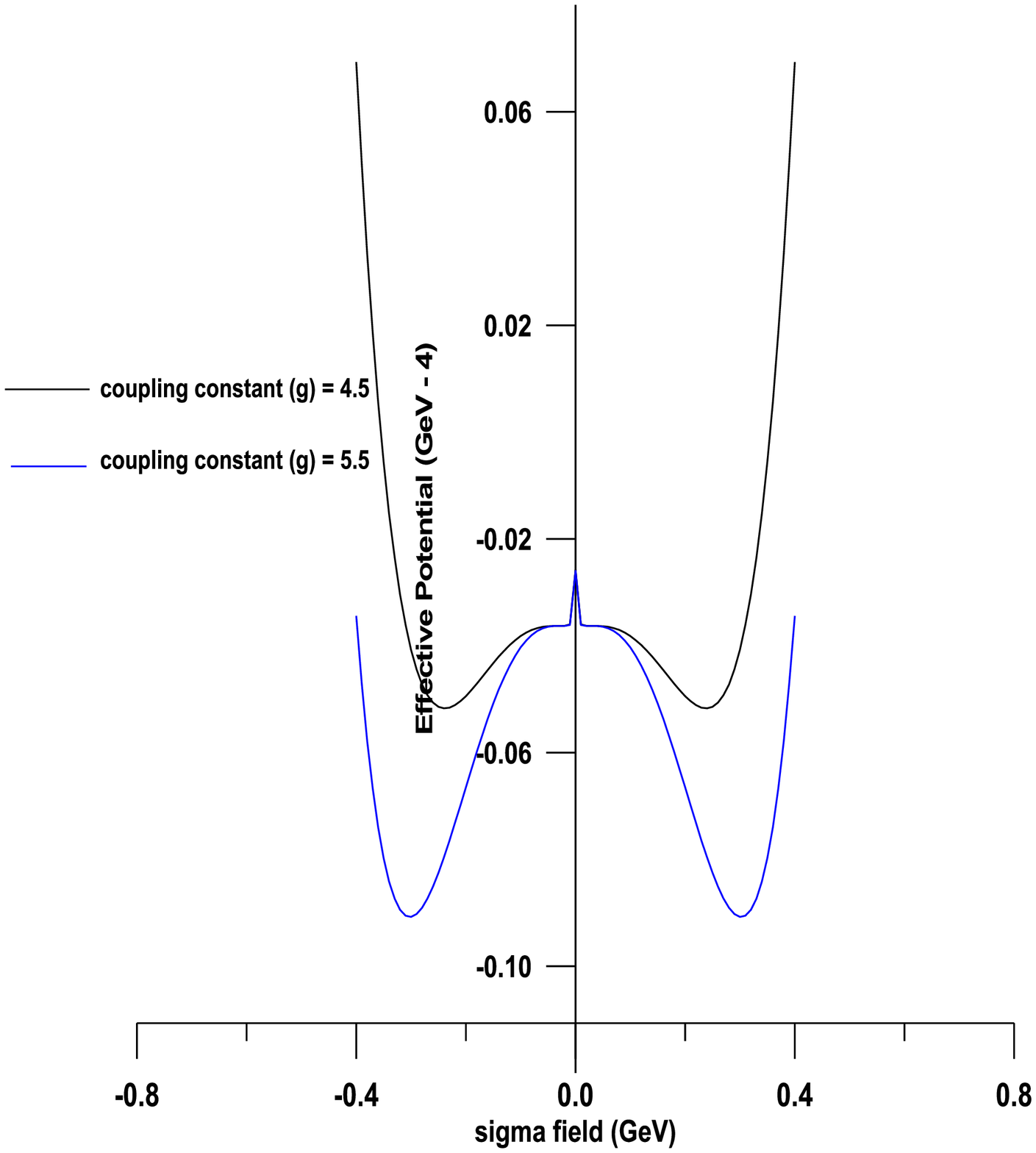}%
\\
\textbf{Fig. 2:} {\small The effective potential is plotted as a function of
sigma field for }$m_{\sigma}=600${\small \ MeV and }$g=4.5${\small \ strong
magnetic field (eB = 0.214 GeV}$^{2})$ {\small in the chiral limit}%
\end{center}

It is important to discuss the effect of free parameters of the model on the
chiral symmetry breaking. So, we select two sets of parameters. First set, the
change of coupling constant with fixed sigma mass as in Fig. 3. \ The
effective potential is plotted as a function of sigma field for different
values of $g$ at dense of magnetic field $\left(  eB=0.214GeV^{2}\right)  $.
The effect of coupling constant $g$ strongly clarify when sigma field
increases. We note that effective potential shifts to higher values by
decreasing $g$ which means that the energy of the potential increases with
decreasing coupling constant $\left(  g\right)  $. In addition, increasing
coupling constant $\left(  g\right)  $ enhances the chiral symmetry breaking
and the potential has two largest minima values in comparison with their
values at $g=4.5$. Therefore, the generate fermionic mass increases. In the
second set of parameters, we fixed the coupling constant $g$ with two values
of sigma mass as in Fig. 4.%

\begin{center}
\includegraphics[
height=4.0413in,
width=4.0413in
]%
{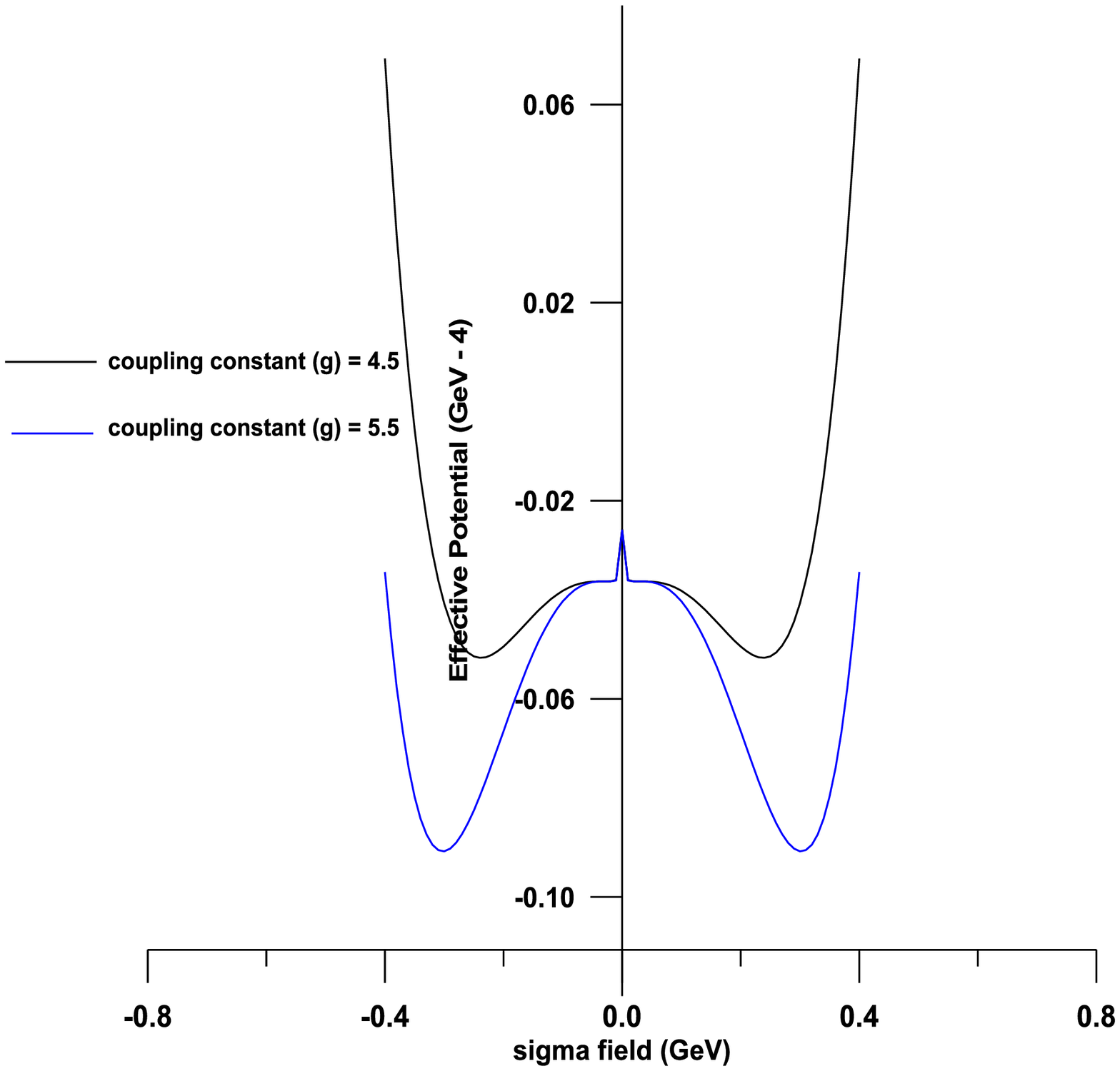}%
\\
\textbf{Fig. 3:} {\small The effective potential is plotted as a function of
sigma field at m}$_{\sigma}=600${\small \ MeV and }$eB=0.214$ {\small GeV}%
$^{2}${\small \ for two values of coupling constant g}$.$%
\end{center}

We note that qualitative features of effective potential are remains
unchanged, in which the effective potential is not sensitive up to $\simeq
\pm175$ MeV and then the potential shifts to higher values by increasing sigma
mass. Also, we note that two minima values of the potential increases with
decreasing sigma mass. Therefore, the decrease in the sigma mass enhances the
chiral symmetry breaking at dense of magnetic field.%

\begin{center}
\includegraphics[
height=4.0413in,
width=4.0413in
]%
{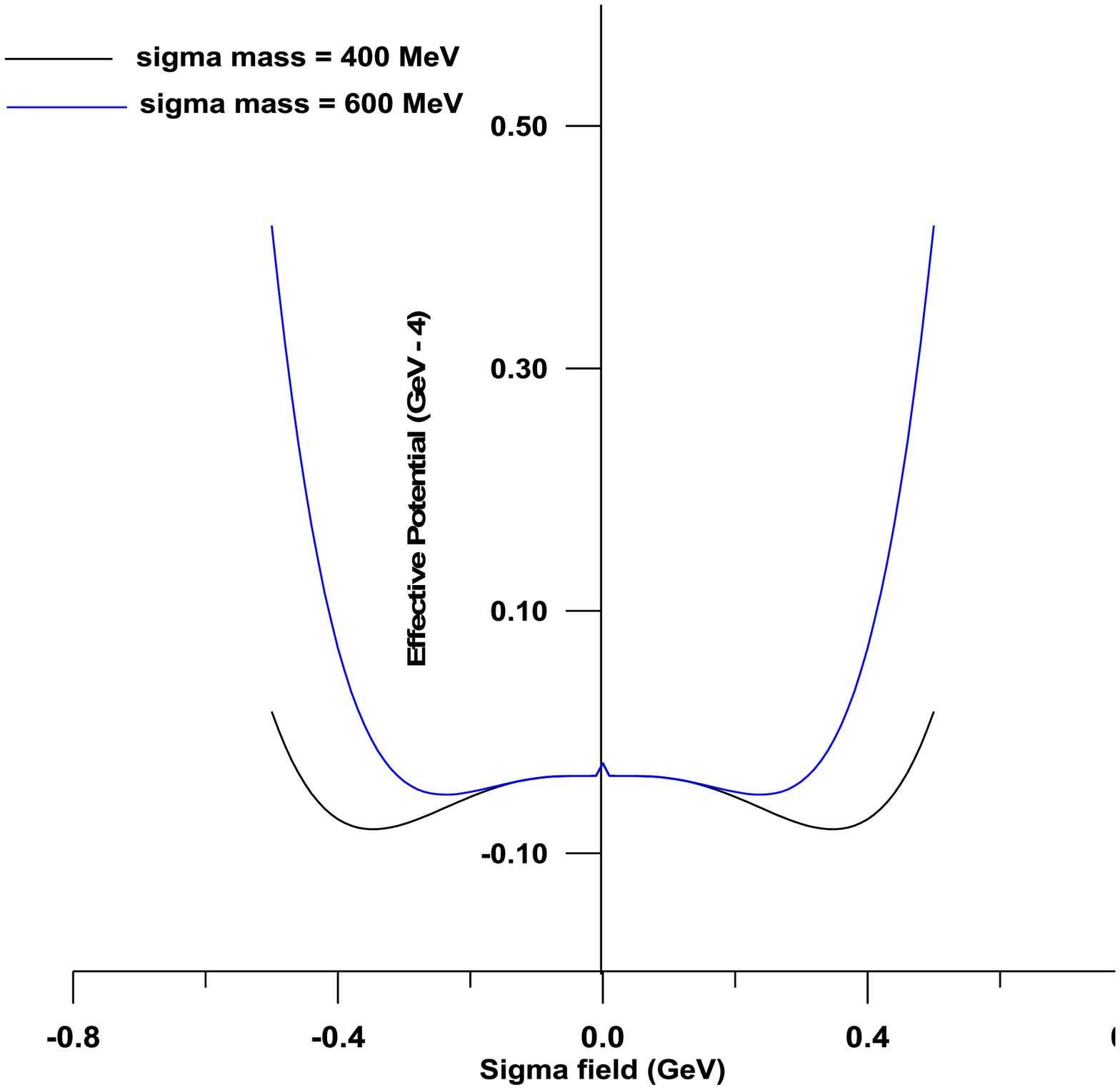}%
\\
\textbf{Fig. 4:} {\small The effective potential is plotted as a function of
sigma field at coupling constant g = 4.5\ and }$eB=0.214$ {\small GeV}$^{2}%
${\small \ for two values of sigma mass}$.$%
\end{center}
%

\begin{center}
\includegraphics[
height=3.5405in,
width=4.0413in
]%
{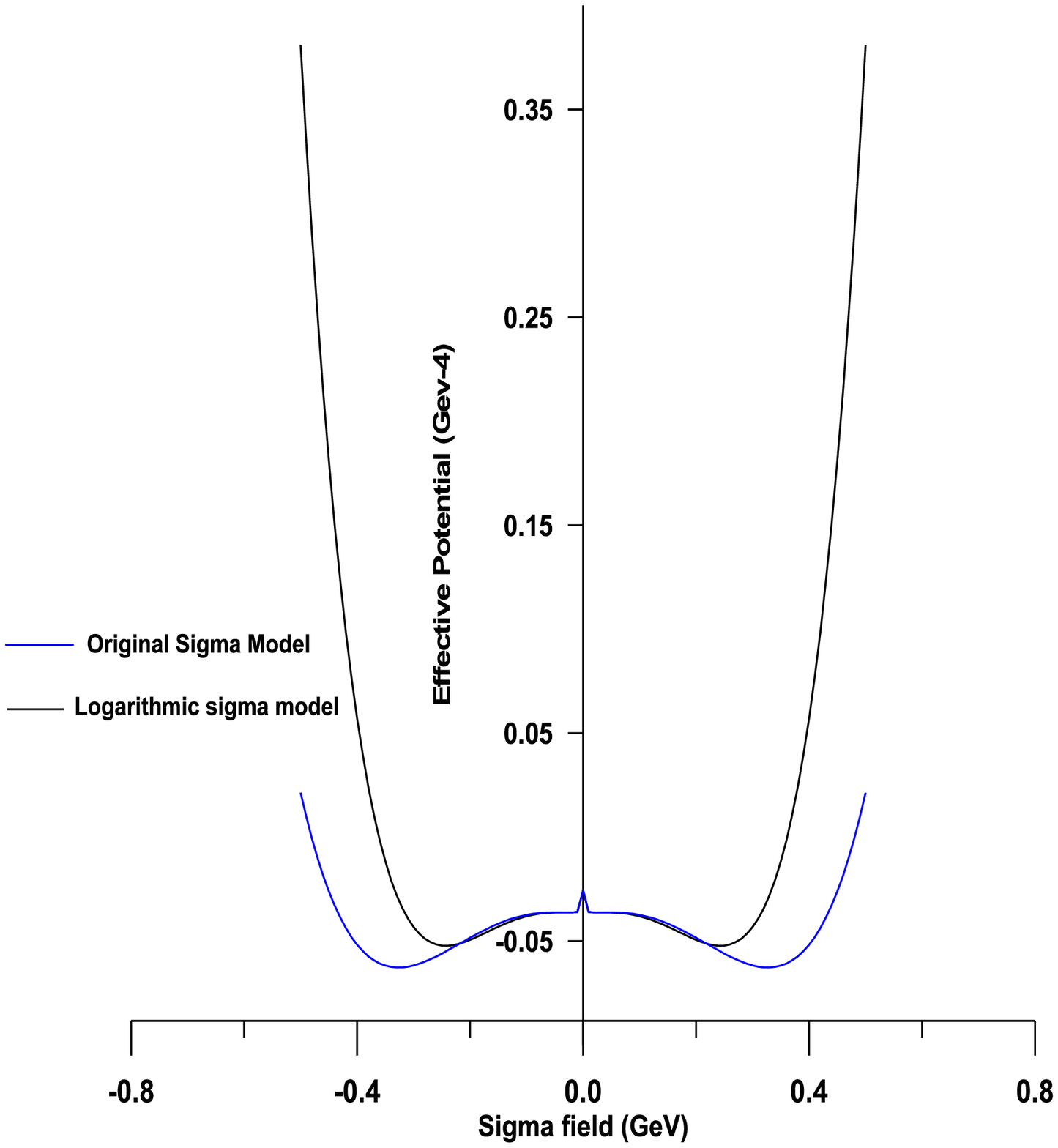}%
\\
\textbf{Fig. 5:} {\small A comparison between the original sigma model and the
logarithmic sigma model at fixed parameters at }$4.5${\small \ \ and
}${\small e}B=0.214${\small \ GeV}$^{2}${\small \ and }$m_{\sigma}%
=600${\small \ MeV .}%
\end{center}

In Fig. 5, we compare between the original sigma model and the logarithmic
sigma model. We fixed all parameters in the two models to show realistic
effect on the chiral symmetry breaking. Qualitative features are similar for
the two models. In addition, we note that the two minima values of logarithmic
potential decreases in comparison with their values in the original sigma model.

\section{Summary and conclusion}

In this work, the effect strange quark flavor on the spontaneous chiral
symmetry breaking in the presence of magnetic field is studied. In addition,
the effect of free parameters of the model is studied in the presence of
strong magnetic field when strange quark flavor is included. A comparison with
original sigma model is presented, showing qualitative agreement with original
sigma model. So, novelty in this work that the effect of strange quark flavor
on the spontaneous chiral symmetry breaking is studied in the framework of the
logarithmic sigma model. Most previous studies have been concentrated on the
original sigma model and NJL model at zero or finite temperature.

We hope to extend the present model to finite temperature and chemical
potential which play an important role for studying properties of the universe
and neutron star.

\section{\textbf{References}}

\begin{enumerate}
\item[1] M. Gell-Mann and M. Levy, Nuono Cinmento \textbf{16}, 705 (1960).

\item[2] M. Birse and M. Banerjee, Phys. Rev. D \textbf{31}, 118 (1985).

\item[3] M. Abu-Shady, Int. J. Mod. Phys. A \textbf{26}, 235 (2011).

\item[4] M. Abu-Shady, Int. J. Theor. Phys. \textbf{48}, 1110 (2009).

\item[5] T. S. T. Aly, M. Rashdan, and M. Abu-Shady, Inter. J. Theor. Phys.
\textbf{45}, 1654 (2006).

\item[6] M. Abu-Shady, Inter. J. Mod. Phys. E \textbf{21}, 1250061 (2012).

\item[7] M. Abu-Shady and M. Soleiman, Phys. of Part and Nucl. Lett
\textbf{10}, 683 (2013)

\item[8] M. Abu-Shady, Inter. J. of Theor. Phys. \textbf{48}, 115 (2009)

\item[9] V. A. Miransky and I. A. Shovkovy, hep-ph%
$\backslash$%
150300732 (2015).

\item[10] S. P. Klevansky, R. H. Lemmer, Phys. Rev. D\textbf{ 39} 3478 (1989).

\item[11] H. Suganuma, T. Tatsumi, Annals Phys. \textbf{208}, 470 (1991)

\item[12] K. G. Klimenko, Z. Phys. C \textbf{54}, 323 (1992).

\item[13] K. G. Klimenko, Theor. Math. Phys. \textbf{90}, 1(1992).

\item[14] Y. Nambu and G. Jona-Lasinio, Phys. Rev. \textbf{122}, 345, (1961);
\textbf{124}, 246 (1961)

\item[15] D. P. Menezes, M. B. Pinto, S. S. Avancini, A. P. Martinez, and C.
Providencia, Phys. Rev. C \textbf{79}, 035807 (2009).

\item[16] D. P. Menezes, M. B. Pinto, S. S. Avancini, and C. Providencia,
Phys. Rev. C \textbf{80}, 065805 (2009).

\item[17] D. Ebert and K. G. Klimenko, Nucl. Phys. A \textbf{728}, 203 (2003).

\item[18] B. Hiller, A. A. Osipov, A. H. Blin, and J. da Providencia, SIGMA
\textbf{4}, 024 (2008).

\item[19] J. O. Andersen and R. Khan, Phys. Rev. D \textbf{85}, 065026 (2012).

\item[20] G. N. Ferrari, A. F. Garcia, and M. B. Pinto, Phys. Rev. D
\textbf{86}, 096005 (2012).

\item[21] M. Abu-Shady, J.  Part. Phys. 1 (1), 58 (2017) hep-ph%
$\backslash$%
1612.02151.

\item[22] M. Abu-Shady, Mod. Phys. Lett. A \textbf{29}, 1450176 (2014).

\item[23] M. Abu-shady and A. K. Abu-Nab, Euro. Phys. J. Plus \textbf{130},
248 (2015).

\item[24] M. Abu-Shady, Appl. Math. Inf. Sci. Lett. \textbf{4}, 5 (2016).

\item[25] M. B. Pinto, D. P. Menezes, and C. Providencia, Phys. Rev. C
\textbf{83}, 065805 (2011).
\end{enumerate}

\end{document}